\documentclass[aps,prl,twocolumn,showpacs,superscriptaddress]{revtex4-1}
\usepackage{graphicx}
\usepackage{amsmath}
\usepackage{amssymb}
\usepackage{colordvi}
\usepackage{mathrsfs}
\usepackage{bm}
\usepackage{verbatim}
\usepackage{dcolumn}
\usepackage{bm}
\usepackage{epsfig}
\usepackage{subfigure}
\usepackage{dsfont}

\usepackage[colorlinks,linkcolor=blue,anchorcolor=blue,citecolor=blue]{hyperref}

\begin{document}
\title{Optical Berry Curvature of Hinge State and Its Detection}
\author{Zheng Liu}
\affiliation{CAS Key Laboratory of Strongly-Coupled Quantum Matter Physics, and Department of Physics, University of Science and Technology of China, Hefei, Anhui 230026, China}
\author{Zhenhua Qiao}
\email[Correspondence author:~~]{qiao@ustc.edu.cn}
\affiliation{CAS Key Laboratory of Strongly-Coupled Quantum Matter Physics, and Department of Physics, University of Science and Technology of China, Hefei, Anhui 230026, China}
\affiliation{ICQD, Hefei National Laboratory for Physical Sciences at Microscale, University of Science and Technology of China, Hefei, Anhui 230026, China}
\author{Yang Gao}
\email[Correspondence author:~~]{ygao87@ustc.edu.cn}
\affiliation{CAS Key Laboratory of Strongly-Coupled Quantum Matter Physics, and Department of Physics, University of Science and Technology of China, Hefei, Anhui 230026, China}
\affiliation{ICQD, Hefei National Laboratory for Physical Sciences at Microscale, University of Science and Technology of China, Hefei, Anhui 230026, China}
\author{Qian Niu}
\affiliation{CAS Key Laboratory of Strongly-Coupled Quantum Matter Physics, and Department of Physics, University of Science and Technology of China, Hefei, Anhui 230026, China}

\date{\today{}}

\begin{abstract}
  We demonstrate that the topological hinge state can possess a nontrivial optical Berry curvature in the nonAbelian formulation of the Berry curvature. It can be readily probed by the circular photogalvanic effect~(CPGE), with the light illuminating a specific hinge, and we refer to it as the hinge CPGE. As a concrete example, we calculate the hinge CPGE in ferromagnetic MnBi$_{2n}$Te$_{3n+1}$, and find that the hinge CPGE peak structure well reflects the optical Berry curvature of hinge states and the optical sum rule captures the optical Berry curvature between the hinge state and the ground state. Thus, the hinge CPGE provides a promising route towards the optical detection of the hinge state geometrical structure.
\end{abstract}

\maketitle

Recent years have witnessed a rapid development of higher-order topological insulators~\cite{2017_Benalcazar,2017_PRB_Benalcazar,2013_Fan,2018_Ezawa,2020_R_Chen,2018_Schindler,2017_Langbehn,2017_Song,2018_Geier,2019_F_Liu,2019_Calugaru,2019_K_Kudo,2020_Ren,2020_Rui_Xing,2019_Trifunovic,2019_Yuanfeng Xu,2019_Sheng,2019_Park,2021_Bing_Liu,2020_Cong Chen,2019_Zhijun Wang,2019_Yue,2019_Pozo,2021_Xie}. Different from the first-order topological insulators, the higher-order topological insulator is characterized by topologically protected states that are at least two dimensional lower than their bulk states. For example, a three-dimensional second-order topological insulator can have topological conducting states localized on the hinges but not on the surfaces~\cite{2018_Geier,2018_Schindler,2017_Langbehn,2017_Song,2019_F_Liu,2019_Calugaru,2019_Trifunovic,2020_Rui_Xing,2019_Zhijun Wang,2019_Yuanfeng Xu,2019_Yue}. Therefore, probing hinge states is indispensable for understanding such second-order topological insulator. By means of scanning tunneling microscope~\cite{2018_Schindler2,2019_Berthold,2022_Nana,2021_Leena}, angle-resolved photoemission spectroscopy~\cite{2021_Noguchi}, and Josephson-interference~\cite{2018_Schindler2,2017_Murani,2020_Choi}, the local density of states is resolved in both the real and momentum space, suggesting the existence of the hinge states. However, besides the characteristic spectrum information, it is still unclear whether the topological nature can also endow hinge states with nontrivial geometrical structures and what consequence they can lead to.

\begin{figure}
	\includegraphics[width=8cm,angle=0]{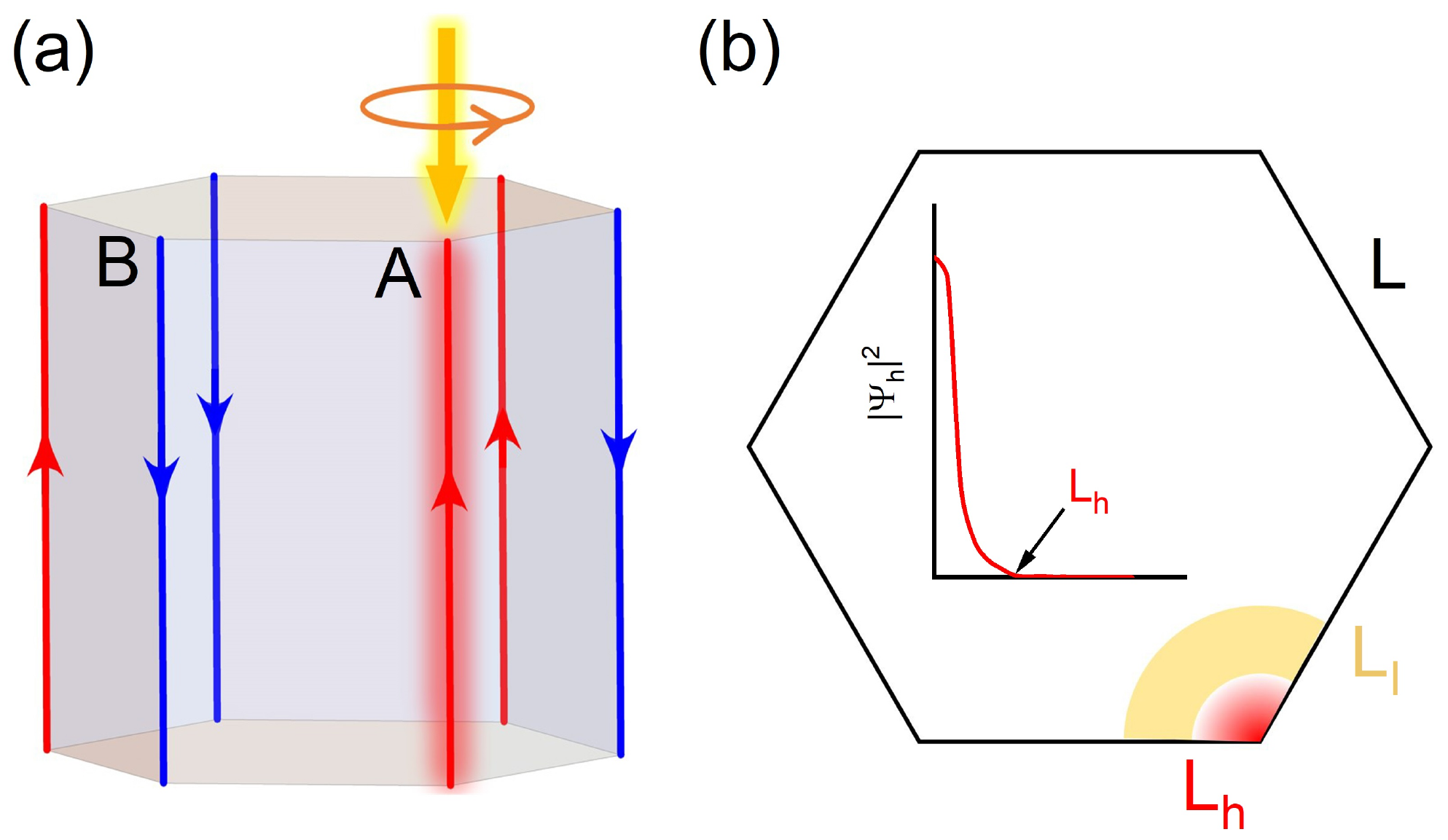}
	\caption{The hinge CPGE~(a) and the corresponding top view of the hexagonal prism~(b). In (a), the topological hinge states along hinge A and B are in red and blue, respectively. The photocurrent around hinge A with illuminating area centered about the hinge is illustrated. In (b), the illuminated region and the localized hinge state are shown explicitly. It is required that $L_h\ll L_I \ll L$.} \label{Fig1}
\end{figure}

In this Letter, we demonstrate that the localized hinge state can have a nonAbelian Berry curvature component, which we refer to as the optical Berry curvature. We then propose a hinge circular photogalvanic effect~(CPGE) as a perfect probe of such optical Berry curvature of the hinge state. The CPGE refers to the part of the photocurrent that switches with the circular polarization of light~\cite{2000_Sipe}, and is an efficient method for capturing the Berry curvature of bulk and surface states~\cite{2011_Hosur,2016_Morimoto,2017_Juan,2020_Alexander,2010_Moore,2009_Deyo,2017_Qiong,2018_Flicker,2020_Holder,2020_Ahn,2018_Suyang,2020_Rees}. By additionally restricting the illuminating area to an appropriate region that fully encapsulates the hinge state spatially yet remains small compared to the sample size, one obtains the hinge CPGE~(see Fig.~\ref{Fig1}). The hinge CPGE involves the optical Berry curvature of the hinge state in a similar fashion with the CPGE and the corresponding sum rule measures the optical Berry curvature between the hinge state and the ground state.

Using MnBi$_{2n}$Te$_{3n+1}$ as a concrete example, we numerically demonstrate the hinge CPGE and its detection of the optical Berry curvature of hinge states. Specifically, we find that the photocurrent of hinge CPGE approaches a steady value as the sample size increases. It also exhibits peak structures due to the optical Berry curvature of the hinge state. 
Furthermore, the optical sum rule indeed well reflects the optical Berry curvature between the hinge state and the ground state.
These properties make the hinge CPGE a promising candidate for detecting the hinge state geometrical structures.

\textit{Optical Berry curvature of hinge states---.}We start with the Berry curvature in extended systems, which plays essential and increasing roles in solid state physics. For example, in the celebrated anomalous Hall effect, one encounters the 
momentum space Berry curvature $\bm \Omega_n=\bm \nabla_{\bm k}\times \bm a_n(\bm k)$, where $n$ is the band index, $\bm a_n(\bm k)=\langle n\bm k|i\bm \partial_{\bm k}|n\bm k\rangle$ is the intraband Berry connection, and $|n\bm k\rangle=e^{-i\bm k\cdot \bm r}|\psi_{n\bm k}\rangle$ is the periodic part of the Bloch function. $\bm \Omega_n$ involves a single band index and is hence Abelian. The origin of such Berry curvature is the restriction of the position operator: $\tilde{r}_i=\hat{P}_n r_i\hat{P}_n$ with $P_n=\sum_{\bm k}|\psi_{n\bm k}\rangle\langle \psi_{n\bm k}|$ being the projection operator. The cross product of $\tilde{\bm r}$ then generates the Abelian Berry curvature: $\bm \Omega^{ab}=-i \tilde{\bm r}\times \tilde{\bm r}=\sum_{\bm k}\bm \Omega_n(\bm k)|\psi_{n\bm k}\rangle\langle \psi_{n\bm k}|$.

In optics, however, the Abelian Berry curvature is not directly measurable and needs to be extended. A good example is the CPGE, where a photocurrent is induced by a circularly polarized light, according to
\begin{align}\label{eq:CPGE1}
\frac{dJ_i}{dt}=\beta_{ij}(\omega)[i\boldsymbol{E}(\omega)\times\boldsymbol{E}^{\ast}(\omega)]_j,
\end{align}
with $\boldsymbol{E}(\omega)$ being the light electric field with frequency $\omega$.
The response function involves a different form of the Berry curvature~~\cite{2000_Sipe}
\begin{eqnarray}\label{eq:CPGE2}
\beta_{ij}&=&-\frac{\pi e^3}{\hbar V}\sum_{\bm k,n,m}f_{nm}\Delta v_{i,nm}(\Omega_n^m)_j\delta(\hbar\omega-\omega_{mn}),
\end{eqnarray}
where $\omega_{mn}=\varepsilon_m-\varepsilon_n$ with $\varepsilon_m$ being the band energy, $f_{nm}=f(\varepsilon_{nk})-f(\varepsilon_{mk})$ with $f(\varepsilon_{nk})$ being the Fermi-Dirac distribution, $\Delta v_{i,nm}=\langle n|\hat{v}_i|n\rangle - \langle m|\hat{v}_i|m\rangle$ with $\hat{v}_i=\frac{1}{\hbar}\frac{\partial \hat{H}}{\partial k_i}$ being the velocity operator, and $\bm \Omega_n^m=-i\langle n\bm k|i\bm \partial|m\bm k\rangle \times \langle m\bm k|i\bm \partial|n\bm k\rangle$ is the geometrical factor.

Interestingly, $\boldsymbol{\Omega}_n^m$ is the difference between the nonAbelian and Abelian Berry curvature. To see this, we define the projection operator onto a pair of bands $n$ and $m$: $\hat{P}=\sum_{\bm k}(|\psi_{n\bm k}\rangle\langle \psi_{n\bm k}|+|\psi_{m\bm k}\rangle\langle \psi_{m\bm k}|)$. At each $\bm k$ point, one readily obtain a $2\times 2$ nonAbelian Berry curvature: $\bm \Omega^{na}=-\frac{i}{2}\hat{e}_\ell \epsilon_{\ell ij} [\hat{P}r_i\hat{P},\hat{P}r_j\hat{P}]$. The geometrical factor can then be expressed as: $\bm \Omega_n^m=\langle \psi_{n\bm k}|\bm \Omega^{na}-\bm \Omega^{ab}|\psi_{n\bm k}\rangle$. The appearance of $\bm \Omega_n^m$ in optical phenomena such as the CPGE relies on two facts: first, the optical transition always relates two bands; secondly, $\bm \Omega_n^m$ is porportional to the oscillator strength of electron-circular-light coupling~\cite{2008_Souza}. Given these features, we thus refer to $\bm \Omega_n^m$ as the optical Berry curvature for a pair of bands. It is invariant under the $U(1)$ gauge transformation of the eigenstate.

More abstractly, $\bm \Omega_n^m$ can also be viewed as the Berry curvature on a $U(2)/[U(1)\times U(1)]$ manifold. At each $\bm k$ point, the projected position operator $\hat{P}\bm r\hat{P}$ is subject to $U(2)$ gauge transfomrations in the subHilbert space spanned by $|n\bm k\rangle$ and $|m\bm k\rangle$. We can then mod out the $U(1)$ gauge freedom for $|n\bm k\rangle$ and $|m\bm k\rangle$ separately, and obtain a $U(2)/[U(1)\times U(1)]$ manifold~\cite{2022_Junyeong}. In such a manifold, the position operator reads: $\bm r^{mod}=\hat{P}\bm r\hat{P}-\hat{P}\bm r\hat{P}|_{U(1)\times U(1)}$ whose diagonal element vanishes. We then have $\bm \Omega_n^m=-i\langle \psi_{n\bm k}|\bm r^{mod}\times \bm r^{mod}|\psi_{n\bm k}\rangle$. We comment that besides the optical Berry curvature, one can also define the optical quantum metric tensor, i.e., $(F_{ij})_{nm}=\frac{1}{2}\langle \psi_{n\bm k}|r_i^{mod} r_j^{mod}+c.c.|\psi_{n\bm k}\rangle$, which is also esstial in nonlinear optical phenomena~\cite{2022_Junyeong}.

The above optical Berry curvature can be readily generalized to be between a pair of states, including localized states such as the topological hinge state. Here we will focus on three-dimensional second order topological insulators. The generalization to other types of localized topological states is straightforward.
Due to the localized nature, the hinge state does not have an Abelian Berry curvature. Assume the hinge state is localized in the $x-O-y$ plane and define the projection operator as: $\hat{P}_h=|\psi_h\rangle\langle \psi_h|$ with $|\psi_h\rangle$ labeling the hinge state. Since $|\psi_h\rangle$ is well localized, $\langle \psi_h|x|\psi_h\rangle$ and $\langle \psi_h|y|\psi_h\rangle$ are well-defined. Then it is straightforward to prove that $[\hat{P}_hx\hat{P}_h,\hat{P}_hy\hat{P}_h]=0$ identically.

In contrast, with the help of a partner state $|\psi_m\rangle$ that can be connected to the hinge state through an optical transition, we can define the optical Berry curvature for a pair of states. To show this, we consider a state pair forming by the hinge state and  $|\psi_m\rangle$ and define $\hat{P}=|\psi_h\rangle\langle \psi_h|+|\psi_{m}\rangle\langle \psi_{m}|$. Using similar arguments as in periodic crystals, we have 
\begin{align}
\Omega_h^m&=-i\langle \psi_h|[\hat{P}x\hat{P},\hat{P}y\hat{P}]|\psi_h\rangle\notag\\
\label{eq_obch} &=-i\langle \psi_h|x|\psi_m\rangle\langle \psi_m|y|\psi_h\rangle-(x\leftrightarrow y)\,.
\end{align}

Such optical Berry curvature can be further generalized by expanding a singe partner state to be a collection of states. This generalization is particularly useful in optics, as the optical sum rule generally relates a continuum of states. To perform such generalization, we use the ground-state projection operator:
\begin{align}
\hat{P}_G=\sum_{m\in occ}|\psi_{m}\rangle\langle \psi_{m}|\,,
\end{align}
where $occ$ stands for the collection of occupied states.
By replacing the partner state projection $|\psi_m\rangle\langle \psi_m|$ with $\hat{P}_G$  in Eq.~\eqref{eq_obch}, we obtain
\begin{align}
\Omega_h^G=-i\sum_{m\in occ}[\langle \psi_h|x|\psi_m\rangle\langle \psi_m|y|\psi_h\rangle]-(x\leftrightarrow y)\,.
\end{align}
This is the optical Berry curvature between the hinge state and the ground state of the topological material. We comment that since $[x,y]=0$ identically, the Berry curvature between hinge state and occupied states differs by a sign from that between hinge state and unoccupied states.

\begin{table*}[t]
	\caption{The scaling properties of various factors in $\beta_{zz}$ due to different optical excitation processes.}
	\begin{ruledtabular}
		\begin{tabular}{ccccccc}
			processes  & $h\leftrightarrow h$ & $h\leftrightarrow s$  &   $h\leftrightarrow b$  &  $s\leftrightarrow s$   &   $s\leftrightarrow b$  &   $b\leftrightarrow b$ \\ \hline    
			$\sum_{a\in I}(\Delta v_z)_{a}$   & $O(1)$   & $O(1)$                     &  $O(1)$                    &  $O(L_I/L)$                  &   $O(L_I/L)$                 &   $O(L_I^2/L^2)$ \\
			$\Omega_n^m$    & $O(1)$  & $O(1/L)$                 &   $O(1/L^2)$               &  $O(L_I^2/L^2)$            &   $O(L_I^2/L^3)$           &   $O(L_I^4/L^4)$ \\
			Multiplicity    & $O(1)$  & $O(L)$                   &   $O(L^2)$                 &  $O(L)$                    &   $O(L^2)$                 &   $O(L^2)$ \\
			$\beta_{zz}^h$    & $O(1)$  & $O(1)$                &   $O(1)$                &  $O(L_I^3/L^2)$         &   $O(L_I^3/L^2)$        &   $O(L_I^6/L^4)$ \\
		\end{tabular}
	\end{ruledtabular}
	\label{table-1}
\end{table*}

{\it Hinge CPGE---.} Strikingly, such an optical Berry curvature can be readily probed using a variant of the CPGE. Without loss of generality, we consider a sample that is finite in the $xy$ plane and periodic along the $z$ direction. We then assume a circularly polarized light propagating along the $z$ direction with an illuminating region labeled by $I$. To probe the hinge state, we further require that the characteristic length scale for the hinge state~($L_h$), illuminating area~($L_I$), and the sample~($L$) satisfy $L_h\ll L_I \ll L$, as shown in Fig.~\ref{Fig1}.  We focus on the induced photocurrent along the same direction and in the same region $I$. The corresponding response coefficient reads~\cite{SM}
\begin{eqnarray}\label{eq:CPGE3}
\beta_{zz}^h(\omega)&=&-\frac{e^3}{2\hbar }\int d k_z\sum_{n,m}^{a\in I}f_{nm}[\langle n|(\hat{v}_z)_a|n\rangle-\langle m|(\hat{v}_z)_a|m\rangle] \notag\\
&&\times(\Omega_n^m)^I\delta(\hbar\omega-\omega_{mn}),
\end{eqnarray}
where $(\hat{v}_z)_a=\{\hat{v}_z,\hat{P}_a\}/2$ projects the velocity on site $a$, and $\hat{P}_a$ is the projection operator with the property $\sum_{a}\hat{P}_a=1$. The geometric factor is given by
\begin{eqnarray}\label{eq:Berry_curvature2}
(\Omega_n^m)^I=\langle n| i[\tilde{x}^I,\tilde{y}^I]|n\rangle,
\end{eqnarray}
where $\tilde{r}_i^I=\hat{P}\hat{P}_I r_i \hat{P}_I\hat{P}$ with the $\hat{P}_I=\sum_{a\in I} \hat{P}_a$ projects onto the illuminating region. It differs from the optical Berry curvature as the additional spatial projection due to the restricted illumination area is needed.
We refer to such photocurrent with restricted illuminating area over one hinge as the hinge CPGE.

Strikingly, $\beta_{zz}^h$ is a thermodynamic property of the sample that only involves the hinge state properties. To prove this, we first note that there are three sets of bands in the sample: the bulk band, the surface band and the hinge band. An incident light with an arbitrary frequency can generally excite electrons within these bands, i.e., there are nine different types of contributions. However, different contributions scale differently with the sample size according to the localized nature of different states. Take the hinge-to-surface or surface-to-hinge process~($h\leftrightarrow s$) as an example. For the site $a$ at the hinge, $\langle a|n\rangle\sim O(1)$ if $|n\rangle$ belongs to the hinge states, while $\langle a|n\rangle\sim O(1/\sqrt{L})$ if $|n\rangle$ belongs to the surface states. The resulting velocity factor in Eq.~\eqref{eq:CPGE3} satisfies: $\langle n|(v_z)_a|n\rangle\sim O(1)$ for hinge states while $\langle n|(v_z)_a|n\rangle\sim O(1/L)$ for surface states. By using similar arguments, we find that the scaling behavior of the geometrical factor: $(\Omega_n^m)^I\sim O(1/L)$.

Besides the obvious velocity and geometrical factor, the summation over $n$ and $m$ brings additional multiplicity. With the increasing sample size, the number of hinge and surface states scale as $O(1)$ and $O(L)$ respectively. Therefore, the summation over the hinge-surface band pair adds a $O(L)$ factor. Putting up all these factors together, we find that the $h\leftrightarrow s$ process scale as $O(1)$ in the thermodynamic limit, i.e., it survives and behaves as a thermodynamic property of the sample.

Using similar arguments, we systematically studied the scaling behavior of the remaining processes. The detail is in the supplementary materials~\cite{SM} and the result is summarized in Table.~\ref{table-1}. It is readily found that in the thermodynamic limit, only the contributions involving the hinge state~(e.g., $h\leftrightarrow h$, $h\leftrightarrow s$ and $h\leftrightarrow b$) survive while all the others vanish. The response fucntion can then be put in a compact form
\begin{align}\label{eq_hcpge}
\beta_{zz}^h(\omega)&=-\frac{ e^3}{2\hbar}\int dk_z\sum_{m}^{n\in H}f_{nm}(v_z)_n \Omega_n^m G_{nm},
\end{align}
where $H$ represents the set of hinge bands near the illuminated hinge, and $G_{nm}=\delta(\hbar\omega-\omega_{mn})-\delta(\hbar\omega+\omega_{mn})$ accounts for the energy conservation. The geometrical factor $\Omega_n^m$ does not involve the real-space projection any more: it reduces to the optical Berry curvature for the hinge state introdued previously. Eq.~\eqref{eq_hcpge} explicitly shows that the hinge CPGE can probe the optical Berry curvature of hinge states.

The optical Berry curvature $\Omega_n^m$ between the hinge state and the ground state can be further extracted using the optical sum rule. To show this, we sum the response function over frequency and define
\begin{align}
\Gamma_h=\int_{0}^{+\infty}\beta_{zz}^h(\omega)d\omega.
\end{align}
At zero temperature, the result reads~\cite{SM}
\begin{align}\label{eq_sum}
\Gamma_h=-\frac{e^3}{2\hbar }\int dk_z \sum_{n\in H}(v_z)_n \Omega_n^G (1-2f_{n}^0),
\end{align}
where $f_n^0$ is the Fermi function at zero temperature. The sum rule in Eq.~\eqref{eq_sum} then involves the optical Berry curvature between the hinge state and the ground state.

The detection of $\Omega_h^G$ can be even more accurate. By taking the derivative with respect to the chemical potential, we have
\begin{align}
\frac{\partial \Gamma_h}{\partial \mu}=\frac{e^3}{\hbar}\sum_{n\in H}^{\varepsilon_n=\mu}\text{sgn}[(v_z)_n] \Omega_n^G (k_F).
\end{align}
Therefore, the variation of the sum rule directly measures the Berry curvature at the Fermi momentum, weighted by the sign of the Fermi velocity. By changing the doping level,  the hinge state Berry curvature can be then mapped across the Brillouin zone.

\begin{figure}
	\includegraphics[width=8.5cm,angle=0]{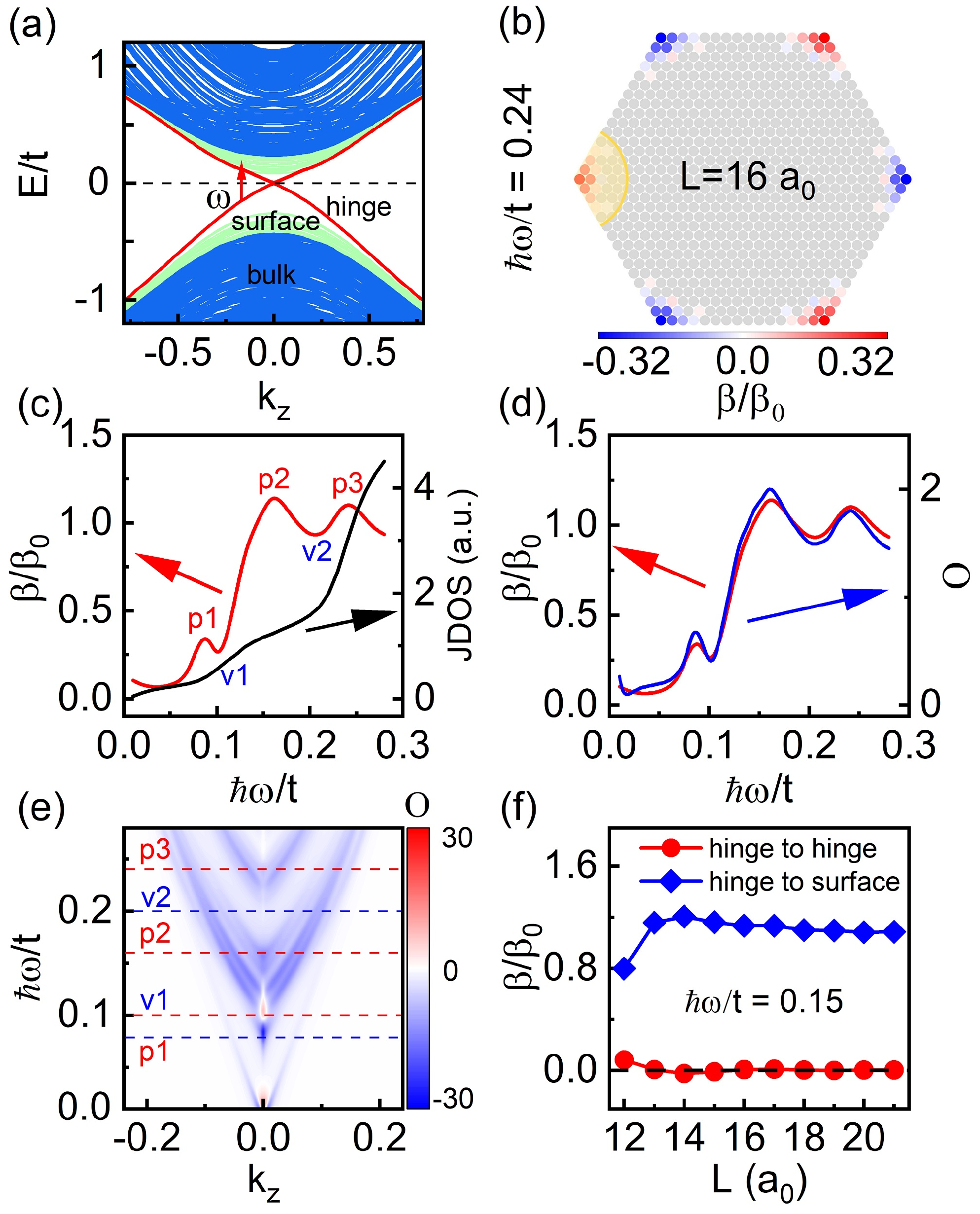}
	\caption{(a) The band structure of the MnBi$_{2n}$Te$_{3n+1}$ with size $L=16$ $a_0$. The hinge, surface, and bulk states are in red, cyan, and blue, respectively. (b) The distribution of $\beta_{zz}(\omega,a)$ in the $xy$ plane at $\hbar\omega/t= 0.24$. (c) The hinge CPGE coefficient and the joint density of states as a function of frequency. $\beta_0=\pi^3e^3/h^2$. (d) $\beta_{zz}^h$ and the average optical Berry curvature $O(\omega)$~(in unit of $a_0/t$). (e) $O(\omega, k_z)$~(in unit of $a_0^2/t$) in the frequency and momentum range.  (f) The hinge to hinge and hinge to surface contribution to $\beta_{zz}^h$ as a function of sample size with $\hbar\omega/t=0.15$. The illuminating region is illustrated in yellow in (b).}
	\label{Fig2}
\end{figure}

\textit{Demonstration in MnBi$_{2n}$Te$_{3n+1}$---.}
As a concrete example, we now demonstrate the optical Berry curvature and hinge CPGE in MnBi$_{2n}$Te$_{3n+1}$. We focus on the ferromagnetic state of MnBi$_{2n}$Te$_{3n+1}$, which is predicted to be a three-dimensional second-order topological insulator~\cite{2020_Rui_Xing}.
In the absence of magnetization, the point group of MnBi$_{2n}$Te$_{3n+1}$ is $D_{3d}$ with the following generators: the spatial inversion $I$, threefold rotation around $z$ axis $C_{3z}$, and twofold rotation around $x$ axis $C_{2x}$. When the magnetization is introduced, the $I$ and $C_{3z}$ are preserved but $C_{2x}$ is replaced by $C_{2x}T$. 
The inversion symmetry forbids the bulk CPGE. For each surface, the inversion symmetry is absent but the $M_xT$ symmetry remains, forbidding the net photocurrent on the surface. Each hinge further breaks $M_xT$ and preserves only the $C_{2x}T$ symmetry, hence permitting the hinge CPGE. The lattice structure, symmetry operations, and the model Hamiltonian are included in the Supplemental Materials~\cite{SM}.

To calculate the hinge CPGE, we consider a hexagonal prism geometry that is periodic in $z$ direction. The side length is set as $L=16~a_0$ with $a_0$ representing the lattice constant. The corresponding energy spectrum is shown in Fig.~\ref{Fig2}(a). We find gapless hinge states between the gapped surface states.

We first calculate the injection current at the hinge as well as over the surface and bulk with the whole sample illuminated~\cite{SM}.
We find that it is nonzero at the hinge but vanishes in the bulk and surface $\beta_{zz}$, consistent with the symmetry analysis. The $\beta_{zz}$ has the same form of Eq.~\ref{eq:CPGE3}, but change the summation of atomic site $a$ to desired region.
The nontrivial injection current at the hinge with the light energy below the surface band gap signifies the existence of the hinge state. To illustrate such an injection current, we plot $\beta_{zz}(\omega)$ at a lattice site $a$ with $\hbar\omega/t=0.24$ in Fig.~\ref{Fig2}(b). One immediately finds that $\beta_{zz}$ is highly localized around the six hinges, with an alternating pattern due to the $D_{3d}$ point group of the whole sample.

We then focus on the left hinge and calculate $\beta_{zz}^h$ corresponding to the hinge CPGE by restricting the illuminating area to be near that hinge. The frequency dependence is shown in Fig.~\ref{Fig2}(c). When the light energy is below 0.08 $t$, the electron can be excited from one hinge state to another; while above 0.08 $t$, the electron can be excited additionally to the surface state. Since more electronic states are involved, one observes a roughly synchronized trend of increase between the hinge CPGE coefficient and the joint density of states, with the latter defined as follows
\begin{eqnarray}\label{eq:JDOS}
{\rm JDOS}=\sum_{m,n}\int\frac{dk_z}{2\pi} f_{nm}\delta(\hbar\omega-\omega_{mn}).
\end{eqnarray}

On top of the synchronized increase trend, the response coefficient shows additional peak structures, which is the manifestation of the optical Berry curvature. Based on Eq.~\eqref{eq:CPGE3}, we can define an average optical Berry curvature associated at a particular frequency $\omega$
\begin{eqnarray}\label{eq:BRBR_BerryBR_Berry}
O(\omega)=\sum_{n,m}\int \frac{dk_z}{2\pi}f_{nm} \Omega_n^m\delta(\hbar\omega-\omega_{mn}).
\end{eqnarray}
It has a clear physical meaning: $\omega^2O(\omega)$ is just the difference of the absorption rate between left and right circularly polarized lights~\cite{2008_Souza,2020_Gao}. We then plot $O(\omega)$ in Fig.~\ref{Fig2}(d). One immediately finds that the coefficient $\beta_{zz}^h$ fully conforms to the peak structure of $O(\omega)$, clearly demonstrating the essential role of the optical Berry curvature.

To study the microscopic origin of the peak structure, we plot the integrant of $O(\omega)$, labeled as $O(\omega,k_z)$ in Fig.~\ref{Fig2}(e). Interestingly, near the $\Gamma$ point, $O(\omega,k_z)$ sharply switches its sign at $\hbar\omega/t=0.09$ and then switches back at $\hbar\omega/t=0.12$, leading to the first peak and valley at around $\hbar\omega/t=0.08$ and $\hbar\omega/t=0.1$. As the frequency further increases, more pairs of bands are involved in the light absorption, reflected by a series of added local extremal at the $\Gamma$ point. The response coefficient thus increases. Between $\hbar\omega/t=0.16$ and $\hbar\omega/t=0.22$, no additional pairs of bands participate and the optical Berry curvature generally decreases away from the band edge, leading to the second valley in the response coefficient.

\begin{figure}
	\includegraphics[width=8.5cm,angle=0]{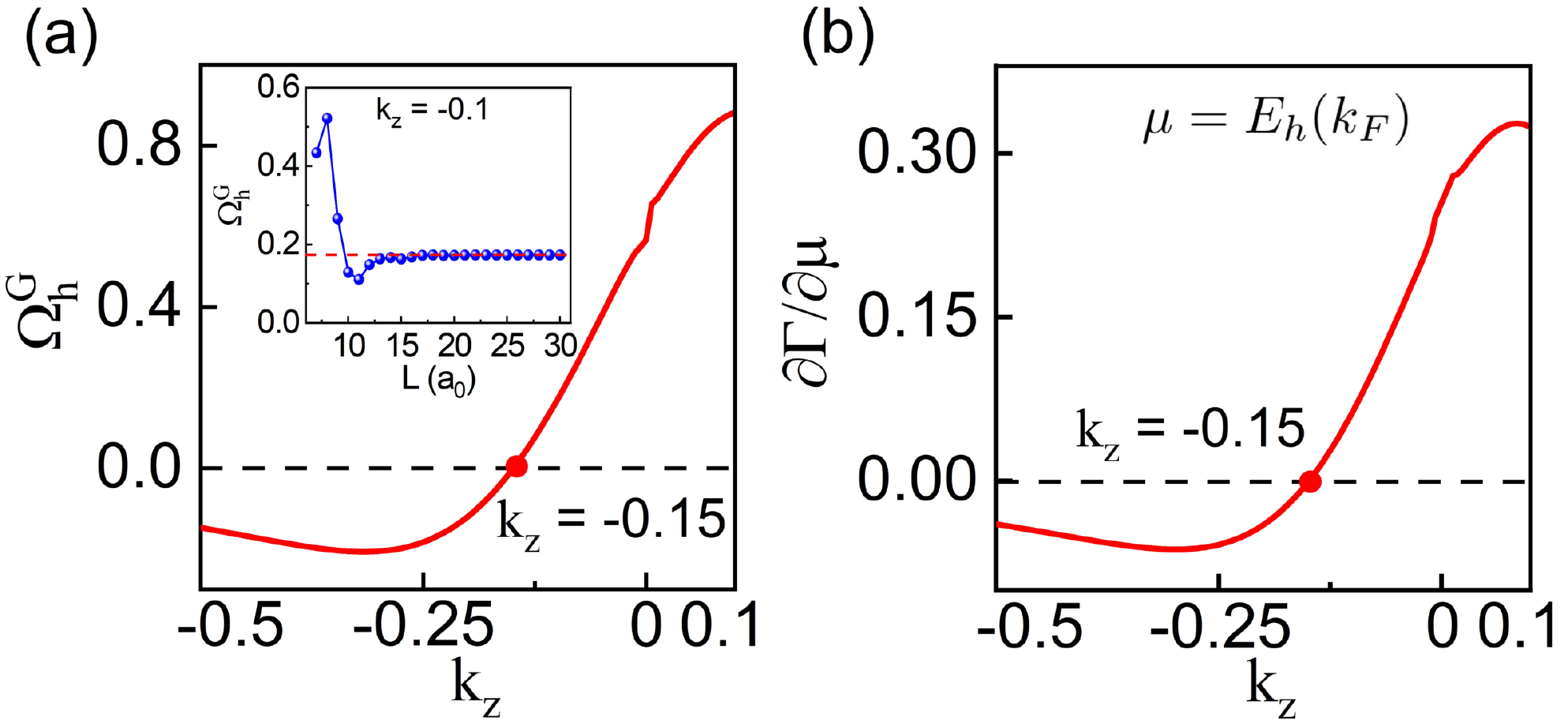}
	\caption{(a) The hinge state Berry curvature~(in unit of $a_0^2$). The sample size is $L=16~a_0$. The inset shows the Berry curvature at $k_z=-0.1$ for different sample sizes. (b) $\partial \Gamma/\partial \mu$ (in unit $a_0^2e^3/\hbar$) as a function of $k_z$. Here, we represent $\partial\Gamma/\partial\mu$ as a function of $k_z$ as each chemical potential corresponds to a unique lattice momentum for the hinge state.}\label{Fig3}
\end{figure}

To clarify the scaling property of hinge CPGE, we choose $\hbar\omega/t=0.15$ and two transition processes contribute to the injection current: $h\leftrightarrow h$ and $h\leftrightarrow s$. In Fig.~\ref{Fig2}(f), we plot these two contributions against the sample size. One observers that, the $h\leftrightarrow h$ contribution gradually vanishes as the two hinges are spatially separated in MnBi$_{2n}$Te$_{3n+1}$. In contrast, the $h\leftrightarrow s$ contribution reaches a steady value, irrelevant with the sample size and consistent with the previous analysis.

Finally, to illustrate the relation between the optical sum rule and the hinge-state Berry curvature. As shown in the inset of Fig.~\ref{Fig3}(a), the optical Berry curvature of hinge state reaches a constant value as the sample become larger. Comparing Figs.~\ref{Fig3}(a) and~\ref{Fig3}(b), we find that they indeed show similar feature confirming the potential usage of the optical sum rule for detecting the hinge state Berry curvature.


\begin{acknowledgments}
 Y. Gao is supported by the National Key R${\rm \&}$D Program under grant Nos. 2022YFA1403502 and Fundamental Research Funds for the Central Universities (Grant No. WK2340000102). Z. Liu and Z. Qiao are supported by the National Natural Science Foundation of China (11974327 and 12004369), Fundamental Research Funds for the Central Universities (WK3510000010, WK2030020032), Anhui Initiative in Quantum Information Technologies (No. AHY170000), and Innovation Program for Quantum Science and Technology (2021ZD0302800). Q. Niu is supported by the National Natural Science Foundation of China (12234017).  The supercomputing service of USTC is gratefully acknowledged.
\end{acknowledgments}

\end{document}